# Laser-excited ultrahigh-resolution photoemission spectroscopy of Na$_x$CoO$_2$·yH$_2$O: Evidence for pseudogap formation

T. Shimojima,[1] T. Yokoya,[1,*] T. Kiss,[1] A. Chainani,[2] S. Shin,[1,2] T. Togashi,[2] C. Chen,[3] S. Watanabe,[1] K. Takada,[4] T. Sasaki,[4] H. Sakurai,[5] and E. Takayama-Muromachi[5]

[1]*Institute for Solid State Physics (ISSP), University of Tokyo, Kashiwa, Chiba 277-8681, Japan*
[2]*The Institute of Physical and Chemical Research (RIKEN),Sayo-gun, Hyogo 679-5143, Japan*
[3]*Beijing Center for Crystal R&D, Chinese Academy of Science Zhongguancun,Beijing 100080, China*
[4]*Advanced Materials Laboratory, National Institute for Materials Science, Tsukuba, Ibaraki 305-0044, Japan.*
[5]*Superconducting Materials Center, National Institute for Materials Science, Tsukuba, Ibaraki 305-0044, Japan.*
[*]*Present address: Japan Synchrotron Radiation Research Institute (JASRI), Spring8, Sayo, Hyogo 679-5198, Japan*

We have studied the temperature-dependent electronic structure near the Fermi level ($E_F$) of the layered cobaltate superconductor, Na$_{0.35}$CoO$_2$·1.3H$_2$O, and related materials, using laser-excited ultrahigh-resolution photoemission spectroscopy. We observe the formation of a pseudogap with an energy scale of ~ 20 meV in Na$_{0.35}$CoO$_2$·1.3H$_2$O and Na$_{0.35}$CoO$_2$·0.7H$_2$O, which is clearly absent in Na$_{0.7}$CoO$_2$. The energy scale of the pseudogap is larger than the expected value for the superconducting gap, suggesting an additional competing order parameter at low temperatures. We discuss implications of the pseudogap in relation to available transport and magnetic susceptibility results.

After the discovery of the first cobalt oxyhydrate superconductor Na$_{0.35}$CoO$_2$·1.3H$_2$O (transition temperature $T_c$ of ~ 4.5 K) by Takada *et al.*,[1] extensive experimental and theoretical studies have been made in order to understand the superconductivity in this system. This is because the conductive two-dimensional CoO$_2$ layers of Na$_{0.35}$CoO$_2$·1.3H$_2$O can be regarded as an electron-doped correlated S=1/2 triangular network of frustrated Co spins,[1] where novel superconductivity emerging from a non Fermi-liquid ground state has been proposed.[2] Theoretically, recent studies on the Co oxide superconductors have proposed novel superconductivity with various types of order parameters depending on the model used.[3] And, several experimental studies have been performed to clarify the symmetry of the superconducting gap.[4] However, agreement on the order parameter has not been reached so far.

In the cuprate high temperature superconductors (high-$T_c$'s), whose electronic structure is often described in terms of a quasi-2 dimensional correlated system, understanding the unusual normal state properties has lead to deeper understanding of the superconductivity. Even for the layered Co oxides, recent studies have indicated a rich phase diagram in the non-hydrated Na$_x$CoO$_2$ system.[5] For superconducting samples, as has been also suggested from theoretical studies,[6-8] importance of well-separated Co$^{3+}$ and Co$^{4+}$ in the superconducting Co oxides has been reported experimentally very recently.[9] Coexistence of superconductivity and ordered states have been discussed in cuprate high-$T_c$ superconductors.[10] In this regard, Na$_{0.35}$CoO$_2$·1.3H$_2$O provides another opportunity to investigate charge and/or spin order competing or cooperating with superconductivity in superconducting oxides.

Thus, it is essential to study the normal-state electronic structure of the superconducting Co oxides using photoemission spectroscopy (PES), which provides direct information on the electronic structure. Very recent PES studies[9] using hard x-rays (escape depth of ~ 50 Å) have provided bulk fundamental physical parameters that describe the new superconductor in terms of, on-site coulomb energy $U_{dd}$, charge transfer energy $\Delta$, and hybridization strength V. More importantly, the double-peak structure in the Co 2p core-level spectra revealed charge disproportionation in Na$_{0.7}$CoO$_2$ and Na$_{0.35}$CoO$_2$·1.3H$_2$O samples, which suggest existence of remnant charge order in the superconducting samples. For the states near the Fermi level ($E_F$), while angle-resolved photoemission spectroscopy (ARPES) studies[11-13] on non hydrated samples showed a large hole pocket around the Γ point in the Brillouin zone, and renormalization on the energy scales of J of ~ 10 meV,[12] no PES study reporting the electronic structure near $E_F$ of superconducting Na$_{0.35}$CoO$_2$·1.3H$_2$O is reported. Therefore, it is extremely important to study superconducting Na$_{0.35}$CoO$_2$·1.3H$_2$O and compare it with that of other related cobalt oxides to investigate the normal-state electronic structure.

For investigating electronic structure of samples having a very delicate surface, PES with a lower photon energy provided by a laser is found to be very powerful,[14] because of the very high energy resolution and expected large escape depth[15]. In this rapid communication, we report temperature-dependent electronic structure near $E_F$ of polycrystalline Na$_{0.7}$CoO$_2$, Na$_{0.35}$CoO$_2$·0.7H$_2$O, and Na$_{0.35}$CoO$_2$·1.3H$_2$O studied with PES using a laser as an excitation source. We succeeded in observing difference in electronic structure between the mother compound and the hydrated samples: absence and presence of a pseudogap. We discuss the pseudogap formation in the light of available transport and magnetic data to elucidate the novel electronic structure of the hydrated superconductor.

Polycrystalline Na$_{0.35}$CoO$_2$·1.3H$_2$O are synthesized from Na$_{0.7}$CoO$_2$ through a chemical oxidation process, by which a part of Na ions is removed and water molecules are intercalated between CoO$_2$ and Na planes, as described in Ref. 1. Magnetic measurements confirmed that the samples measured here have $T_c$ of 4.5 K. PES under an ultrahigh vacuum condition requires special caution because superconducting Na$_{0.35}$CoO$_2$·1.3H$_2$O has a tendency to become non superconducting Na$_{0.35}$CoO$_2$·0.7H$_2$O due to loss of water molecules. Therefore, we carefully handled hydrated samples as follows. We first covered the samples with silver paste and mounted them on copper substrates to prevent loss of water molecules under vacuum. Then, the prepared samples were cooled to ~180 K and are fractured *in-situ*. Immediately after the fracturing, they are transferred to a measurement chamber and measured without warming up above 180 K. We chose the temperature of 180 K because no loss of water molecules occurs in $Na_{0.35}CoO_2 \cdot 1.3H_2O$ below 250 K.[16] In actual practice, the pressure did not change during the fracturing and measuring the hydrated samples. Therefore, we believe that the water molecule content is unchanged even after fracturing and measuring them under vacuum.

All PES measurements were performed on a spectrometer built using a GAMMADATA-SCIENTA R-4000 electron analyzer and an ultraviolet laser (hν = 6.994 eV). The best energy resolution is 360 μeV.[14] However, the energy resolution for all measurements was set to 6.5 meV in order to scan wide energy range and to get reasonable count rate. Samples are cooled using a flowing liquid He refrigerator with improved thermal shieldings. Sample temperatures were measured using a silicon-diode sensor mounted below the samples. The base pressure of the measurement chamber was better than $2 \times 10^{-11}$ Torr. The spectra were all reproducible during the measurement of 2 hours. $E_F$ of samples was referenced to that of gold film evaporated onto the sample substrate. Its accuracy is estimated to be better than ± 0.1 meV. The advantage of the ultraviolet laser is the larger escape depth as expected from the literature (~ 200 Å) of emitted electrons[15], enabling a probe of bulk electronic states.

Figure 1 shows the PES spectra near $E_F$ of (a) $Na_{0.7}CoO_2$, (b) $Na_{0.35}CoO_2 \cdot 0.7H_2O$, and (c) $Na_{0.35}CoO_2 \cdot 1.3H_2O$ measured as a function of temperature from 3.5 K to 150 K. The spectra were normalized with the total intensity integrated from 64 meV to - 56 meV binding energy. The similar intensity of measured spectra around 60 meV of each sample confirms the validity of the normalization we used. All the spectra show systematic temperature dependence with a clear Fermi edge structure at lower temperatures. However, we also found that there is a notable difference in the spectral shape near $E_F$ at 3.5 K between $Na_{0.7}CoO_2$ and the two hydrated samples. While the spectrum of $Na_{0.7}CoO_2$ has a nearly constant intensity with a smooth reduction close to $E_F$, the spectra of $Na_{0.35}CoO_2 \cdot 0.7H_2O$ and $Na_{0.35}CoO_2 \cdot 1.3H_2O$ show an additional decrease within 20 meV binding energy of $E_F$. Close look at the temperature-dependent intensity at $E_F$ also shows a small but an important difference; while the intensity at $E_F$ of $Na_{0.7}CoO_2$ remains constant, the intensity at $E_F$ of the two hydrated samples decreases with lowering temperature. This is in sharp contrast to the temperature dependence of a normal metal, like gold, where all the temperature dependent spectra intersect with each other at $E_F$ with the 1/2 value of the density of states (DOS) at $E_F$. This is clear from the insets where we show an enlarged view of intensities at $E_F$ for the three samples. Thus, the raw spectra itself indicates that the two hydrated samples exhibit an unusual normal-state electronic structure near $E_F$.

In order to make the temperature-dependent change in

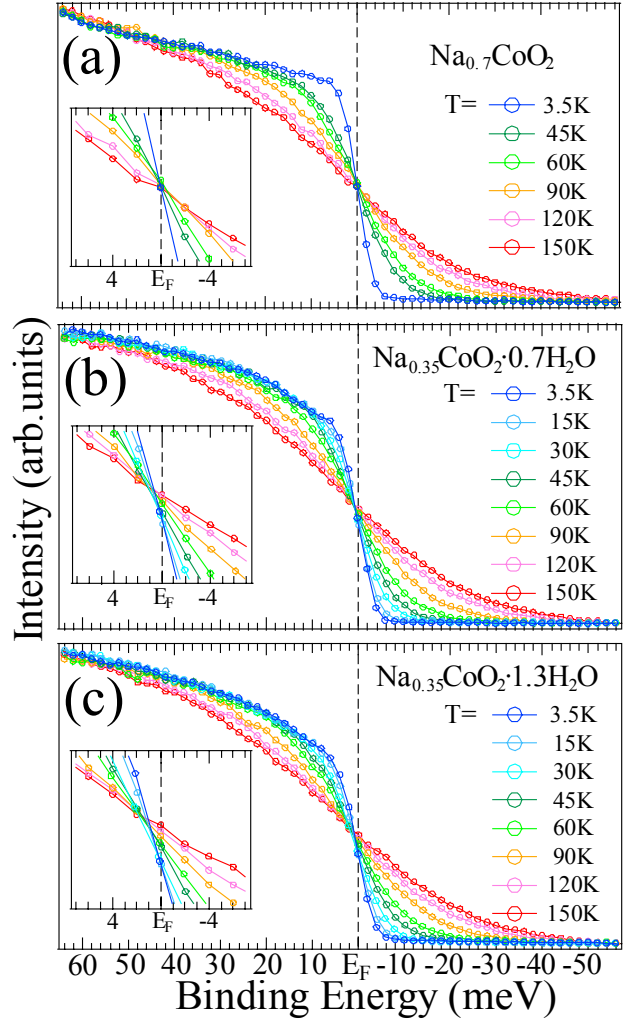

Fig.1. Temperature-dependent spectra of (a) $Na_{0.7}CoO_2$, (b) $Na_{0.35}CoO_2 \cdot 0.7H_2O$, and (c) $Na_{0.35}CoO_2 \cdot 1.3H_2O$ measured with 6.994 eV photon energy. Insets are the enlargement of the near $E_F$ region, which highlight the difference in temperature dependent change at $E_F$ between $Na_{0.7}CoO_2$ and the hydrated samples.

DOS clearer, we have performed an analysis to extract the effect of the Fermi-Dirac (FD) function from the raw data as follows. The raw spectrum at each temperature was divided by the corresponding FD function convoluted with a Gaussian with the instrumental resolution. To see the temperature-dependent spectral changes, the divided spectra at all temperatures were further divided with the smoothed spectrum at 150 K, as shown in Figs. 2 (a)-(c).[17] We also plotted in Fig.3 the temperature-dependent intensity at $E_F$ (averaged from –2meV to 2 meV binding energy of the raw spectra in Fig. 1) normalized with that at 150 K of each sample. In Figs. 2 (b) and (c), the normalized DOS of the hydrated samples within ~ 20 meV of $E_F$ are gradually suppressed with decreasing temperature. The reduced spectral weight around $E_F$ seems to be compensated by a slight increase beyond 20 meV binding energy, indicating that spectral weight is transferred from the near-$E_F$ region to the higher-binding energy region. Compared with the marked change in the hydrated samples, the temperature-dependent change in $Na_{0.7}CoO_2$ is smaller, though at lower temperatures the normalized DOS shows a broad hump centered around 30 meV and a small upturn near $E_F$. Absence of the pseudogap in $Na_{0.7}CoO_2$ is consistent with the angle-integrated PES studies on $Na_{0.6}CoO_2$.[18] These qualitative differences in the spectra between $Na_{0.7}CoO_2$ and the hydrated samples are also seen in the temperature dependent intensity at $E_F$ as shown in Fig. 3. Relatively flat intensities in $Na_{0.7}CoO_2$ and a decrease of the intensity with decreasing temperatures in the hydrated samples. It should be noted that the qualitative difference is not due to difference in inhomogeneity of samples, as we know that the quality is better for the hydrated samples because of the narrower core levels.[9] This is also known from tunneling experiments which indicted less inhomogeneity in $Na_{0.35}CoO_2 \cdot 1.3H_2O$ than in $Na_{0.7}CoO_2$ and which also observed a pseudogap.[19] Thus, we conclude that appearance of the pseudogap in the hydrated samples and absence of the pseudogap in $Na_{0.7}CoO_2$ reflects the intrinsic difference in electronic structure, highlighting unusual electronic states realized in the hydrate samples.

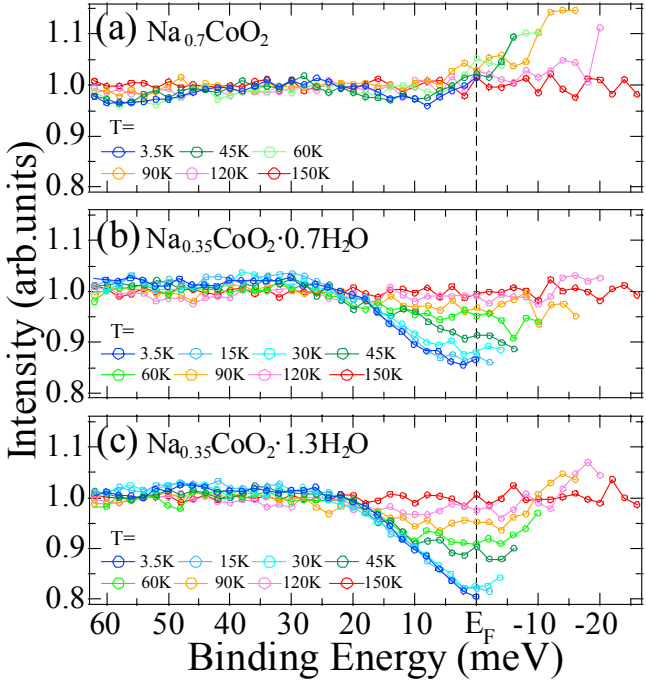

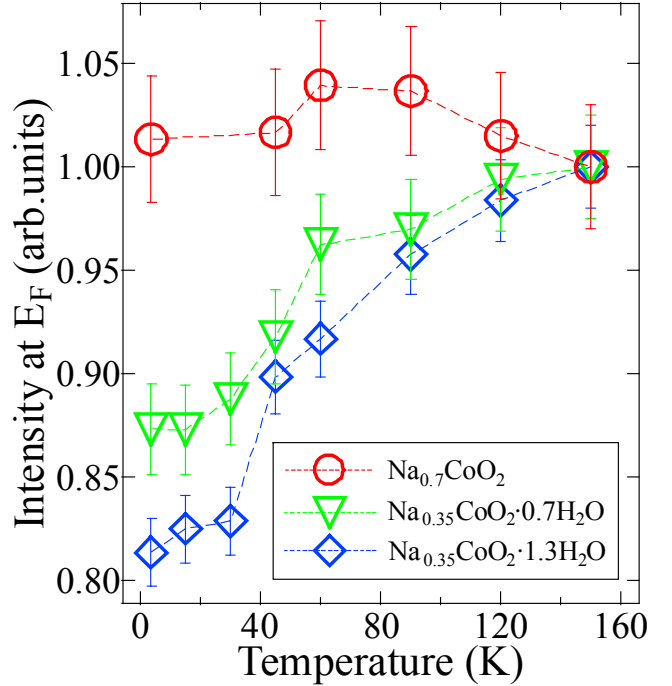

Fig. 2. Temperature-dependent normalized DOS of (a) $Na_{0.7}CoO_2$, (b) $Na_{0.35}CoO_2 \cdot 0.7H_2O$, and (c) $Na_{0.35}CoO_2 \cdot 1.3H_2O$. Note that clear formation of pseudogaps at lower temperatures in the hydrated samples in (b) and (c).

Fig.3. Temperature-dependent intensity at $E_F$ (averaged from –2 meV to 2 meV binding energy) of $Na_{0.7}CoO_2$ (circles), $Na_{0.35}CoO_2 \cdot 0.7H_2O$ (triangles) and $Na_{0.35}CoO_2 \cdot 1.3H_2O$ (diamonds). All intensities were normalized with that of 150 K.

More recently, evidence of temperature dependent pseudogap formation for non hydrated layered cobalt oxides $Na_{0.25}CoO_2$ and $Na_{0.5}CoO_2$ has been reported from optical spectroscopy.[20] The energy scales of half of the pseudogap measured from the optical spectroscopy are 6-25 meV, depending on the Na concentration, consistent with the value obtained from the present study. However, the smaller reduction in the intensities (10-20% reduction compared with that of 150 K) in the PES results compared with the optical results indicates that the pseudogap formation in hydrated samples is partly suppressed due to the difference in concentration of Na and/or inclusion of the water molecules, suggesting parts of Fermi surface (FS) are gapped. This anticipation may have something to do with the resistivity data[21,22] not showing an anomaly around 40 K, where intensity at $E_F$ of $Na_{0.35}CoO_2 \cdot 1.3H_2O$ shows a marked decrease (Fig. 3). One possibility is that the pseudogap opens on the bands having little contribution to the transport properties, as described below. According to the band calculation,[7] $Na_{0.33}CoO_2$ has two types of bands across $E_F$ derived from mainly Co 3d electrons. One has a dominant $a_{1g}$ character providing light carriers and forming a large cylindrical FS centered around the Γ-point. The other has a primarily $e_g'$-like character providing heavy carriers and forming a small pocket FS along the Γ-K direction. The $e_g'$-like band mainly contributes to the density of state at $E_F$. From these aspects, one may say that the pseudogap forms on the small pockets but the large FS plays a dominant role for the transport. Moreover, recent theoretical studies have predicted that there is strong nesting tendencies for the small pockets that can induces charge or spin density waves (CDW or SDW).[7] It may be likely that the pseudogap opens as a precursor of the CDW or SDW. Absence of pseudogap in the DOS of $Na_{0.7}CoO_2$ may be related to the disappearance of the small pockets in $Na_{0.7}CoO_2$ due to the rigid band filling of electrons. Indeed, ARPES have reported the absence of the small pockets in $Na_xCoO_2$ (x = 0.7).[12] To check the presence of the small pocket, which plays a crucial role for the novel superconductivity, ARPES of the hydrated samples is desired.

In relation to the superconductivity, the energy scale of the pseudogap (20 meV) is larger than the expected SC gap size (~ 1 meV as expected from the $T_c$ and the known mean field relation of $2\Delta/k_BT_c = 3.54$, where $\Delta$ is the SC gap value, $k_B$ the Boltzmann constant). This indicates that the pseudogap is different from the superconducting gap, suggesting possible competing ground states at low temperatures. A large energy scale pseudogap has been reported for the cuprate high-$T_c$ superconductors,[23,24] where the pseudogap formation was found to be related to a development of magnetic correlations. On the other hand, a temperature-dependent pseudogap has been observed in $Ba_{0.67}K_{0.33}BiO_3$,[25] where the energy scale of the pseudogap correlated with the energy of the breathing mode phonon. In the layered Co oxides, the energy scales of ~ 20 meV is comparable to the nearest neighboring magnetic coupling constant $J$ = 10-20 meV as predicted for $U_{dd}$ ~ 5 eV,[7] which has been also concluded from ARPES measurements.[12,13] This $J$ value is smaller than the highest phonon energies.[20] It is interesting to note that temperature-dependent magnetic susceptibility shows deviations from a Pauli-like susceptibility beginning at the same temperature at which we see pseudogap formation or intensity reduction at $E_F$ more for $Na_{0.35}CoO_2 \cdot 1.3H_2O$. The comparative study of temperature dependent susceptibility of $Na_{0.35}CoO_2 \cdot 0.7H_2O$ and $Na_{0.35}CoO_2 \cdot 1.3H_2O$ has shown an enlarged magnetic susceptibility in $Na_{0.35}CoO_2 \cdot 1.3H_2O$ compared to $Na_{0.35}CoO_2 \cdot 0.7H_2O$.[27] The NMR studies have also reported an existence of ferromagnetic fluctuation at lower temperatures.[28] The additional suppression in DOS at $E_F$ (Figs 2 and 3) in $Na_{0.35}CoO_2 \cdot 1.3H_2O$ compared to $Na_{0.35}CoO_2 \cdot 0.7H_2O$ also suggests further stabilization of the competing order. While these observed interesting physical properties in hydrated and non-hydrated $Na_xCoO_2$ needs more systematic studies to be understood, the present results showing a pseudogap may give an insight into the physics underlying the unconventional superconductivity observed in the hydrated layered cobalt oxide superconductors.

In conclusion, we have studied the electronic structure near $E_F$ of $Na_{0.35}CoO_2 \cdot 1.3H_2O$ and related cobalt oxides using laser-excited ultrahigh-resolution PES. The obtained results show that, while the temperature-dependent spectra of the mother material shows no gap structure near $E_F$, those of hydrated samples show pseudogap with an energy scale of 20 meV in the occupied part with larger depletion at $E_F$ in $Na_{0.35}CoO_2 \cdot 1.3H_2O$ compared in $Na_{0.35}CoO_2 \cdot 0.7H_2O$. The energy scale of the pesudogap (20 meV) is larger than the expected superconducting gap size (~ 1 meV), which indicates that the pseudogap is a necessary but not sufficient condition for the superconductivity. These results suggest the superconductivity in the hydrated Co oxide occurring close to some ordered phases.

We thank S. Maekawa, W. Koshibae and Y. Yanase for their valuable discussion. We also thank F. Kanetaka for his help in PES measurements. This study was partially supported by CREST of Japan Science and Technology Agency (JST), and by Grants-in-Aid for Scientific Research (B) (16340111) and Young Scientists (A) (14702010) from Japan Society for the Promotion of Science.